%% file: HIP41378f.tex
\documentclass[fleqn,usenatbib]{mnras}

\usepackage{newtxtext,newtxmath}

\usepackage[T1]{fontenc}

\DeclareRobustCommand{\VAN}[3]{#2}
\let\VANthebibliography\thebibliography
\def\thebibliography{\DeclareRobustCommand{\VAN}[3]{##3}\VANthebibliography}

\input{commands.tex}
\input{properties.tex}
\usepackage{color}

\usepackage{xcolor}
\usepackage{hyperref}
\newcommand{\orc}{\includegraphics[height=\fontcharht\font`A]{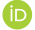}}
\newcommand{\orcid}[1]{\href{https://orcid.org/#1}{\orc}}
\usepackage{graphicx}	
\usepackage{amsmath}	

\title[TTV for \Nplanet]{A transit timing variation observed for the long-period extremely low density exoplanet \Nplanet }

\author[E. M. Bryant et al.]{
\parbox{\textwidth}{
Edward M.~Bryant\orcid{0000-0001-7904-4441},$^{1,2}$\thanks{E-mail: edward.bryant@warwick.ac.uk}
Daniel~Bayliss,$^{1,2}$
Alexandre Santerne,$^{3}$
Peter J.~Wheatley,$^{1,2}$
Valerio Nascimbeni,$^{4}$
Elsa Ducrot,$^{5}$
Artem Burdanov,$^{5,6}$
Jack S. Acton,$^{7}$
Douglas R. Alves, $^{8}$\orcid{0000-0002-5619-2502}
David R. Anderson,\orcid{0000-0001-7416-7522}$^{1,2}$
David J. Armstrong,$^{1,2}$
Supachai Awiphan,$^{9}$
Benjamin F. Cooke,$^{1,2}$
Matthew R. Burleigh,\orcid{0000-0003-0684-7803}$^{7}$ 
Sarah L. Casewell,$^{7}$
Laetitia Delrez,$^{5,10,11}$
Brice-Olivier Demory,$^{12}$
Philipp Eigm\"uller, $^{13}$
Akihiko Fukui,$^{14,15}$
Tianjun Gan\orcid{0000-0002-4503-9705},$^{16}$
Samuel Gill,$^{1,2}$
Michael Gillon,$^{29}$
Michael R. Goad,$^{7}$
Thiam-Guan Tan\orcid{0000-0001-5603-6895},$^{17}$
Maximilian N. G{\"u}nther\orcid{0000-0002-3164-9086},$^{18}$\thanks{Juan Carlos Torres Fellow}
Bronwen Hardee\orcid{0000-0002-1767-1148},$^{19}$
Beth A. Henderson,$^{7}$
Emmanuel Jehin,$^{10}$
James S. Jenkins,$^{8,20}$
Molly Kosiarek\orcid{0000-0002-6115-4359},$^{21}$\thanks{NSF Graduate Research Fellow}
Monika Lendl,$^{11}$\orcid{0000-0001-9699-1459}
Maximiliano Moyano,$^{22}$\orcid{0000-0002-7927-9555}
Catriona A. Murray\orcid{0000-0001-8504-5862},$^{23}$
Norio Narita,$^{15,24,25,26}$
Prajwal Niraula\orcid{0000-0002-8052-3893},$^{6}$
Caroline E.~Odden,$^{27}$
Enric Palle,$^{15,28}$
Hannu Parviainen,$^{15,28}$
Peter P. Pedersen\orcid{0000-0002-5220-609X},$^{23}$
Francisco J. Pozuelos\orcid{0000-0003-1572-7707},$^{29,10}$
Benjamin V.\ Rackham\orcid{0000-0002-3627-1676},$^{6}$\thanks{51 Pegasi b Fellow}
Daniel Sebastian\orcid{0000-0002-2214-9258},$^{30}$ 
Chris Stockdale\orcid{0000-0003-2163-1437},$^{31}$
Rosanna H. Tilbrook,$^{7}$
Samantha J. Thompson\orcid{0000-0002-8039-194X},$^{23}$ 
Amaury H.M.J. Triaud\orcid{0000-0002-5510-8751},$^{30}$ 
St\'{e}phane~Udry,$^{11}$
Jose I. Vines, $^{8}$\orcid{0000-0002-1896-2377}
Richard G.~West,$^{1,2}$
Julien de Wit,$^{6}$
}
\\
$^{1}$Dept.\ of Physics, University of Warwick, Gibbet Hill Road, Coventry CV4 7AL, UK\\
$^{2}$Centre for Exoplanets and Habitability, University of Warwick, Gibbet Hill Road, Coventry CV4 7AL, UK\\
$^{3}$Aix Marseille Univ, CNRS, CNES, LAM, Marseille, France\\
$^{4}$INAF - Osservatorio Astronomico di Padova, Vicolo dell'Osservatorio 5,
I-35122 Padova, Italy\\
$^{5}$Astrobiology Research Unit, Universit{\'e} de Li{\`e}ge, All{\'e}e du 6 Ao{\^u}t 19C, B-4000 Li{\`e}ge, Belgium\\
$^{6}$Department of Earth, Atmospheric and Planetary Science, Massachusetts Institute of Technology, 77 Massachusetts Avenue, Cambridge, MA 02139, USA\\
$^{7}$School of Physics and Astronomy, University of Leicester, Leicester, LE1 7RH, UK\\
$^{8}$Departamento de Astronom\'ia, Universidad de Chile, Casilla 36-D, Santiago, Chile\\
$^{9}$National Astronomical Research Institute of Thailand, 260 Moo 4, Donkaew, Mae Rim, Chiang Mai, 50180, Thailand\\
$^{10}$Space Sciences, Technologies and Astrophysics Research (STAR) Institute, Universit\'e de Li\`ege, 19C All\`ee du 6 Ao\^ut, 4000 Li\`ege, Belgium\\
$^{11}$Observatoire Astronomique de l'Universit\'e de Gen\`eve, Chemin Pegasi 51, Versoix, Switzerland\\
$^{12}$University of Bern, Center for Space and Habitability, Gesellschaftsstrasse 6, 3012 Bern, Switzerland\\
$^{13}$Institute of Planetary Research, German Aerospace Center, Rutherfordstrasse 2, 12489 Berlin, Germany\\
$^{14}$Department of Earth and Planetary Science, Graduate School of Science, The University of Tokyo, 7-3-1 Hongo, Bunkyo-ku, Tokyo 113-0033, Japan\\
$^{15}$Instituto de Astrof\'{i}sica de Canarias (IAC), 38205 La Laguna, Tenerife, Spain\\
$^{16}$Department of Astronomy and Tsinghua Centre for Astrophysics, Tsinghua University, Beijing 100084, China\\
$^{17}$Perth Exoplanet Survey Telescope (PEST)\\
$^{18}$Department of Physics, and Kavli Institute for Astrophysics and Space Research, Massachusetts Institute of Technology, Cambridge, MA 02139, USA\\
$^{19}$Department of Earth and Planetary Sciences, University of California, Santa Cruz, CA 95064, USA\\
$^{20}$Centro de Astrof\'isica y Tecnolog\'ias Afines (CATA), Casilla 36-D, Santiago, Chile\\
$^{21}$Department of Astronomy and Astrophysics, University of California, Santa Cruz, CA 95064, USA\\
$^{22}$Instituto de Astronom\'ia, Universidad Cat\'{o}lica del Norte, Angamos 0610, 1270709, Antofagasta, Chile\\
$^{23}$Astrophysics Group, Cavendish Laboratory, J.J. Thomson Avenue, Cambridge CB3 0HE, UK\\
$^{24}$Komaba Institute for Science, The University of Tokyo, 3-8-1 Komaba, Meguro, Tokyo 153-8902, Japan\\
$^{25}$JST, PRESTO, 3-8-1 Komaba, Meguro, Tokyo 153-8902, Japan\\
$^{26}$Astrobiology Center, 2-21-1 Osawa, Mitaka, Tokyo 181-8588, Japan\\
$^{27}$Phillips Academy Observatory, Phillips Academy, Andover MA USA\\
$^{28}$Departamento de Astrof\'isica, Universidad de La Laguna (ULL), 38206, La Laguna, Tenerife, Spain\\
$^{29}$Astrobiology Research Unit, Universit\'e de Li\`ege, 19C All\`ee du 6 Ao\^ut, 4000 Li\`ege, Belgium\\
$^{30}$School of Physics \& Astronomy, University of Birmingham, Edgbaston, Birmingham B15 2TT, United Kingdom\\
$^{31}$Hazelwood Observatory\\}

\date{Accepted 2021 April 2. Received 2021 March 31; in original form 2021 March 10}

\pubyear{2021}

\begin{document}
\label{firstpage}
\pagerange{\pageref{firstpage}--\pageref{lastpage}}
\maketitle

\begin{abstract}
\Nplanet\ is a temperate \Nradius\rearth\ planet with period of \Nperiodshort\,days and an extremely low density of \Ndensity\gccc.  It transits the bright star \Nstar\ (V=\NVmagshort), making it an exciting target for atmospheric characterization including transmission spectroscopy. \Nstar\ was monitored photometrically between the dates of 2019 November 19 and November 28. We detected a transit of \Nplanet\ with NGTS, just the third transit ever detected for this planet, which confirms the orbital period. This is also the first ground-based detection of a transit of \Nplanet. Additional ground-based photometry was also obtained and used to constrain the time of the transit. The transit was measured to occur \Ndtcshort\,hours earlier than predicted.  We use an analytic transit timing variation (TTV) model to show the observed TTV can be explained by interactions between \Nplanete\ and \Nplanet. Using our TTV model, we predict the epochs of future transits of \Nplanet, with derived transit centres of \tcfour\,=\,\Ntcfour\ (May 2021) and \tcfive\,=\,\Ntcfive\ (Nov 2022).
\end{abstract}




\section{Introduction}
The bright (V=\NVmagshort) F-type star \Nstar\ is known to host five transiting planets \citep{Vanderburg16hip41378}, based on photometric data taken during campaign C5 of the \ktwo\ mission \citep{Howell14K2}. The outer most planet known in the system, \Nplanet\, only transited once during these observations.  A second transit of \Nplanet\ was observed during \ktwo\ campaign C18 and showed it to have a maximum possible orbital period of \NPmax\,days \citep{Becker19hip41378}.  Extensive spectroscopic monitoring with HARPS
by \citet{Santerne19hip41378} revealed the true period of \Nplanet\ to be \Nperiodshort\,days - half the possible maximum.

With a mass of \mpl$=$\Nmass\,\mearth\ and a radius of \rpl$=$\Nradius\,\rearth, the low density of \Nplanet, combined with its cool temperature,
makes it a particularly interesting target for atmospheric characterization. Using the system parameters from \citet{Santerne19hip41378} we calculate a Transmission Spectroscopy Metric (TSM) for \Nplanet\ of 343 \citep{kempton2018tsm}, which shows \Nplanet\ to be very well suited to transmission spectroscopy follow-up.  For these observations, a precise orbital ephemeris is needed. Despite the high quality \ktwo\ data, with only two recorded transits there was previously no information regarding any transit timing variations (TTVs) of \Nplanet. In a multi-planet system, a planet with a long orbital period such as \Nplanet\ may experience large transit timing variations (TTVs) \citep[eg.][]{Agol05ttvs, KawaharaMasuda19longperiodKepler}. Since \Nplanet\ resides in near 2:1 and 3:2 mean motion resonances with planet e and d respectively, we may expect TTVs on the order of hours to days.

\begin{figure}
    \centering
    \includegraphics[width=\columnwidth]{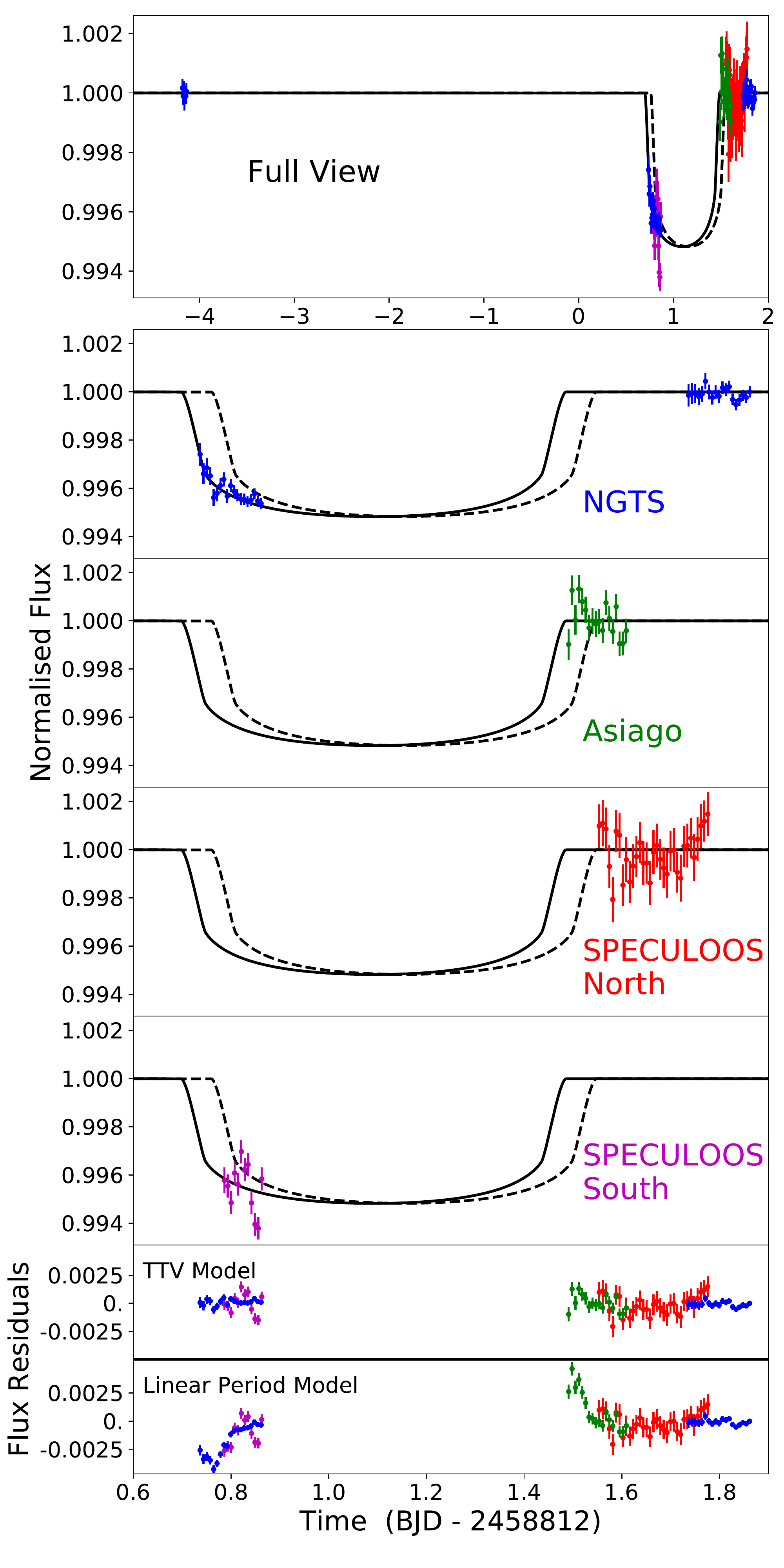}
    \caption{Ground-based relative time-series photometry of \Nstar. All the photometric data included in the analysis are shown in the top panel. The individual datasets are presented in the following panels: NGTS (blue; second panel) showing a zoom in on the transit of \Nplanet, ASIAGO (green; third panel), SNO (red; fourth panel) and SSO (magenta; fifth panel).  All data are binned to 10\,minutes.  The predicted transit model of \Nplanet\ from the K2 and HARPS data is given by the dashed black line, and the model from our sampling process is given by the solid black line. The residuals to the TTV model from our fitting are presented in the sixth panel and the residuals to the predicted linear period model from the two prior transits are presented in the bottom panel. We note that the colours used to denote the individual datasets are consistent across each panel. Note that we plot the detrended NGTS data and the relative flux offsets between the photometry obtained at different facilities were fit as free parameters during the modelling set out in Section~\ref{sec:results}.}
    \label{fig:phot}
\end{figure}

Ground-based photometric observations of the \Nstar\ system are needed to measure any TTVs to place additional constraints on the masses and orbital periods of planets in the \Nstar\ system. Monitoring the TTVs also has the potential to reveal other exoplanets in the system, and additionally will help predict future transit events for atmospheric characterisation. However, for a star with a magnitude of V=\NVmagshort\,  the field-of-view of most ground-based photometric facilities will often not include enough comparison stars of a similar brightness \citep{collins18}. In fact, for \Nstar, the nearest star with a GAIA G magnitude difference of less than 0.5\,mag is 17\,arcmin away \citep{GAIA_DR2}.

\section{Photometric Observations}\label{sec:obs_phot}
\subsection{NGTS}\label{sub:obs_ngts}
The Next Generation Transit Survey \citep[NGTS; ][]{Wheatley18NGTS} is a photometric facility situated at ESO's Paranal Observatory in Chile. It consists of twelve fully-robotic telescopes with 20\,cm diameter apertures and wide fields-of-view of 8 square degrees. The dominant photometric noise sources in NGTS bright star light curves have been shown to be Gaussian and uncorrelated between the individual telescopes \citep{bryant20multicam}. Combined with the wide field-of-view, this allows NGTS to use simultaneous observations with multiple telescopes to achieve high precision light curves of bright stars \citep{smith20multicam, bryant20multicam}.

\Nstar\ was observed with NGTS on the nights UT 2019 Nov 19, 24 and 25. On all nights, \Nstar\ was monitored using eleven NGTS telescopes, each using the custom NGTS filter (520-890\,nm) and an exposure time of 10\,seconds. Across all nights, a total of 23841 images were taken. The target was above an altitude of 30$^{\circ}$ for all the observations, and the observing conditions were good. 

The NGTS data were reduced using a custom aperture photometry pipeline, which utilises the SEP Python library \citep{sextractor1996, Barbary16sep}. The pipeline uses the GAIA DR2 catalogue \citep{GAIA, GAIA_DR2} to automatically identify comparison stars that are similar in magnitude, colour and CCD position to \Nstar\ \citep[for more details see][]{bryant20multicam}. Comparison stars which a shown to display variability or high levels of photometric scatter are excluded from the reduction. The NGTS photometry is plotted in Figure~\ref{fig:phot}.

\subsection{SPECULOOS}\label{sub:speculoos}
\Nstar\ was observed by \spec\ \citep[Search for habitable Planets EClipsing ULtra-cOOl Stars;][]{Burdanov2018,Gillon2018,Delrez18SPECULOOS,Sebastian2021}. It was observed from the SPECULOOS Southern Observatory (SSO) at Paranal, Chile on UT 2019 Nov 24 and from the SPECULOOS Northern Observatory (SNO) at the Teide Observatory on Tenerife, Spain on UT 2019 Nov 25. The SSO observations consists of 342 images obtained using the $r'$ filter, an exposure time of 12\,seconds, and a defocus of 300 steps to avoid saturation. SNO obtained 254 images using also the $r'$ filter, with an exposure time of 12\,seconds and a defocus of 300 steps. For each night of observation, the automatic SSO Pipeline \citep{Murray2020} was used to reduce images and extract precise photometry. The SSO Pipeline is built upon the casutools software \citep{Irwin2004} and performs automated differential photometry to mitigate ground-based systematics, including a correction for time-varying telluric water vapour. The SNO and SSO data are plotted in Figure~\ref{fig:phot}.

\subsection{Asiago Telescope}\label{sub:asiago}
\Nstar\ was observed on the night UT 2019 Nov 25 using the 67/92-cm Schmidt telescope based at the Asiago Observatory in northern Italy, operated by the Italian National Institute of Astrophysics (INAF). A
total of 301 images were obtained through the Sloan $r$ filter under a good sky quality, using a constant exposure time of 30 seconds and defocusing the camera to minimize pixel-to-pixel systematic errors. The
light curve was extracted by using a custom version of the STARSKY code, a pipeline to perform high-precision differential photometry originally developed for the TASTE project \citep{nascimbeni2011a}. The unbinned
RMS is 2.0 mmag, the scatter being much larger at the beginning of the series due to the much higher airmass (1.97 for the first frame). All the time stamps were converted to the BJD-TDB system \citep{eastman2010}. The Asiago data, binned over a 10-min time scale (rms: 825 ppm) are plotted in Figure 1 (green points). The Asiago data are plotted in Figure~\ref{fig:phot}.

\subsection{Photometric Transit Detection}\label{sub:transdetect}
The NGTS data shows a clear flux decrease which is consistent to the predicted transit depth and near to the predicted transit time. This flux decrease is clearly seen in the top two panels of Figure~\ref{fig:phot}. We note that the NGTS photometry is of sufficient precision that an ingress occurring at the time predicted from the K2 and HARPS data would be robustly detected. The SSO and Asiago data are consistent with the change in transit time seen by NGTS, ruling out an on-time ingress and egress respectively. The SNO data rule out a late transit. With photometric coverage on three separate nights surrounding the transit event, we are confident that the flux offset seen in the NGTS data is a real signal. However the relative flux offsets between the data from the different facilities, including the offset of the SSO data, were fit independently as free parameters during the modelling detailed in Section~\ref{sec:results}. The combination of these four datasets gives us confidence that the transit occurred earlier than the prediction from the linear orbital ephemeris derived from the K2 and HARPS data \citep{Santerne19hip41378}. This is the third transit of this planet to be detected and the first from the ground, with the other two being detected from data obtained by the \ktwo\ mission. The detection of this transit confirms the \Nperiodshort\,day period predicted for this planet from K2 and HARPS data \citep{Santerne19hip41378}.

\subsection{Additional Ground Based Photometry}
Additional photometry for \Nstar\ was also obtained with various other ground-based facilities: PEST, LCO (CTIO), PROMPT-8 (CTIO), MUSCAT2 at Teide Observatory, Lick Observatory, Hazelwood Observatory, and Phillip's Academy Observatory. Unfortunately, due to poor observing conditions or a lack of suitable comparison stars, none of the data from these facilities had a photometric precision better than a threshold of 2\,ppt-per-10\,mins. As such we do not include these datasets in our analysis.

\section{Analysis}\label{sec:results}
We jointly modelled the photometry from NGTS, ASIAGO, SSO and SNO in order to determine the time of transit centre for the November 2019 transit, \tcthree.  Note we use the notation $T_{C, N}$ to refer to the time of transit centre for transit epoch $N$. The first transit of \Nplanet\ in \ktwo\ C5 is taken to occur at \tczero.  We generated a transit light curve template based on the parameters from \citet{Vanderburg16hip41378}, and using the period derived by \citet{Santerne19hip41378}. We performed an MCMC sampling process to model the ground-based photometric data. The free parameters used in the modelling were \tcthree, the limb darkening coefficients, the detrending coefficients for the NGTS data, and relative flux offset factors for the other datasets. Note that each NGTS single-telescope lightcurve was detrended against airmass independently from the other telescopes, and a single linear trend with time is applied to the entire NGTS multi-telescope light curve. We used a quadratic limb darkening profile, constraining the coefficients based on theoretically derived limb-darkening coefficients for the NGTS filter. The r' filter used for the SSO observations has a large amount of overlap with the wavelength range of the NGTS filter, and so the limb-darkening coefficients derived for the NGTS filter provide a good approximation for the SSO data. The precision of the SSO data also means that including an additional set of limb-darkening coefficients for the SSO data does not have an effect on the derived value of \tcthree. Therefore, we use the theoretical values of \NLDA\ and \NLDB, which were computed from the \citet{claretbloemen2011LDcoeffs} tables, as Gaussian priors for the model for all datasets.

The sampling was performed using the Ensemble Sampler from \texttt{emcee} \citep{foremanmackey13}. We ran \nwalkers\ chains for \nsample\ steps, following a burn-in phase of \nburnin\ steps. The number of effective samples for each parameter ranged from 600000 to 800000, with the number of effective samples for \tcthree\ was 698539.39784769. From this analysis, we obtained a transit time of \tcthree\,=\,\Ntcthree, which is \Ndtc\,hours earlier than predicted. 
The best fit model from this analysis is shown in Fig.~\ref{fig:phot} and the derived transit time variation is shown in Fig.~\ref{fig:ttv}.

\section{TTV Analysis}\label{sec:ttv_analysis}
\begin{figure}
    \centering
    \includegraphics[width=\columnwidth]{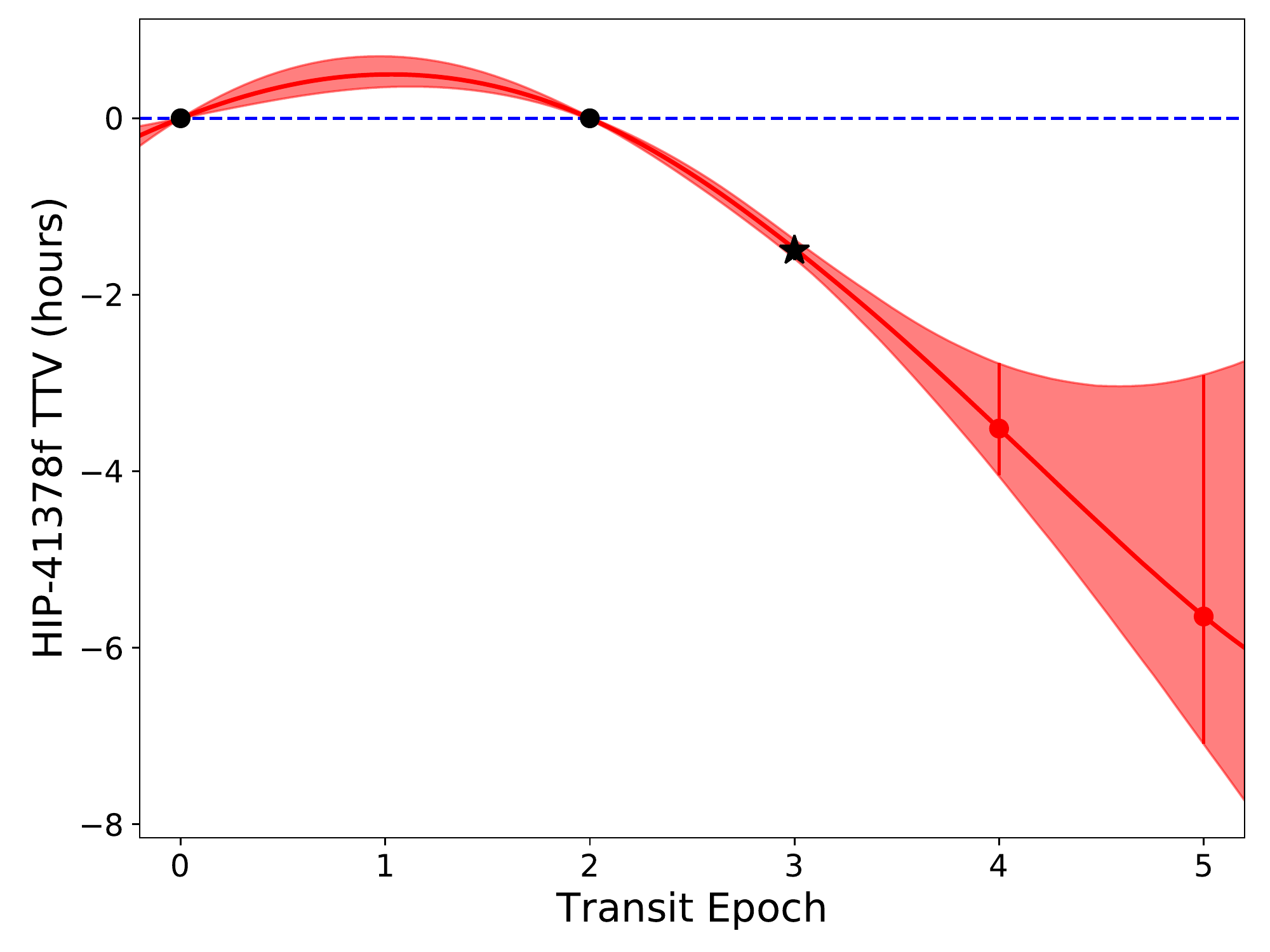}
    \caption{Transit times for \Nplanet.  The black points give the $T_C$ values for the two transits of \Nplanet\ observed by \ktwo. The black star gives the new \tcthree\ determined in this work. The 1$\sigma$ errorbars on these values are plotted but are two small to be visible. The red points and errorbars give the predicted transit times for \tcfour\ (May 2021) and \tcfive\ (Nov 2022) and their respective 1$\sigma$ uncertainties. The solid red line gives the median TTV signal from the TTV analysis (see Section~\ref{sec:ttv_analysis}), and the red shaded area gives the 1$\sigma$ uncertainty on this signal. The TTV values are relative to the linear ephemeris based on the period from \citet{Santerne19hip41378} of $T_0 =$ \Ntcvan\ and $P_f = 542.07975$~days, and this linear ephemeris is shown by the blue dashed line.}
    \label{fig:ttv}
\end{figure}
Due to its anomalously low density, \Nplanet\ is an attractive target for atmospheric characterisation studies. As such, knowing the precise ephemerides for future transits is of high importance. Therefore, we use analytical TTV formulae from \citet{lithwick12ttv} to predict the times of the upcoming transits of \Nplanet, in particular the next two transits in May 2021 and Nov 2022.

Of the planets already known in the system, the TTV signal of \Nplanet\ will be most affected by \Nplanetd\ and \Nplanete. These two planets have masses of $M_d < 4.6$\mearth\ and $M_e = 12\pm5$\mearth\ \citep{Santerne19hip41378}, and reside near to 2:1 and 3:2 mean motion resonances with \Nplanet, respectively. The amplitude of the TTVs of \Nplanet\ as a result of its interactions with \Nplanetd\ and \Nplanete\ will depend strongly on $M_d$ and $M_e$ respectively. The interactions with \Nplanete\ are likely to dominate the TTV signal of \Nplanet\ since $M_e > M_d$ \citep{Santerne19hip41378} and thus we consider solely the TTVs due to interactions between \Nplanet\ \& e for the rest of this analysis. We note that neither \Nplanetd\ nor \Nplanete\ were robustly detected by \citet{Santerne19hip41378} in HARPS radial velocity measurements, and so they are only able to place loose constraints on the masses and periods of these planets. The constraints on the period of \Nplanete\ are derived from a combination of transit analyses and asteroseismic stellar constraints \citep{Santerne19hip41378, Lund19hip41378asteroseismology}. Future monitoring of the TTVs of \Nplanet\ will allow us to place tighter independent constraints on the masses and periods of \Nplanetd\ and \Nplanete.

We use the \citet{lithwick12ttv} analytic formulae for TTV signals for planets in a near mean motion resonance to obtain the following equations for the amplitude, $V$, and super-period, $P_{\rm ttv}$, for the TTV signal of \Nplanet\ as a result of its interactions with \Nplanete. These are
\begin{equation}
    V = P_f \frac{M_{\rm pl, e}}{3\pi \Delta M_*}\left(-G + \frac{3}{2\Delta}\left(F\,e_e + G\,e_f\right)\right),
\end{equation}
\begin{equation}
    P_{\rm ttv} = \frac{P_f}{3|\Delta|} ,
\end{equation}
where $P_f$ is the mean linear orbital period of \Nplanet, $M_{\rm pl, e}$ and $M_*$ are the masses of \Nplanete\ and \Nstar, $e_e$ and $e_f$ are the orbital eccentricities of \Nplanete\ and \Nplanet\ respectively, and $F$ and $G$ are coefficients given in Table 3 of \citet{lithwick12ttv}. The $\Delta$ parameter is the normalized distance to resonance
\begin{equation}
    \Delta = \frac{P_f}{P_e}\frac{2}{3} - 1,
\end{equation}
where $P_e = 369 \pm 10$\,days is the mean linear orbital period of \Nplanete, . Using the planetary parameters given in Tab.1 of \citet{Santerne19hip41378}, we calculate values of $V = 1.3667$\,days, and $P_{\rm ttv} = 8757.577$\,days. We note that the expected TTV signal from \Nplanete\  alone is enough to account for the TTVs observed in Section~\ref{sec:obs_phot}.  Additional planets in the system are not required to explain the observations (nor do the observations rule them out).

With just three transit times, we cannot use TTV modelling to place any constraints on the masses and orbital eccentricities of the planets in the \Nstar\ system. Neither can we robustly refine the period of \Nplanete. However, we can combine the analytical TTV signals derived in \citet{lithwick12ttv} with our knowledge of \Nplanet\ and e to predict the times of the next transits of \Nplanet. We do this using an MCMC sampling method, again using \texttt{emcee} Ensemble Sampler. The TTV model we use for this sampling is of the form
\begin{equation}
    T_{C, N} = T_0 + P_f N + V \sin\left( \frac{2\pi P_f}{P_{\rm ttv}} (N + \phi)\right)
\end{equation}
where $T_{C, N}$ is the time of transit centre of the $N^{\rm th}$ transit, $T_0$ is a reference epoch and $N$ is the the transit epoch, such that $T_0 + P_f N$ is the mean linear ephemeris of \Nplanet, and $\phi$ is the phase shift of the TTV signal. The free parameters in this sampling are $T_0$, $P_f$, $\phi$, $M_{\rm pl, e}$, $P_e$, $e_e$ and $e_f$. For the following parameters: $M_{\rm pl, e}$, $P_e$, $e_e$ and $e_f$ we impose Gaussian priors using the posterior values from \citet{Santerne19hip41378}. In this way, we explore the parameter space of TTV signals which can explain the transit times we observe for \Nplanet\ and are also physically plausible. For the sampling, we run 40 walkers each for 200000 steps as a burn-in, and then a following 50000 steps to explore the parameter space. The resultant predicted TTV signal from this sampling is given in Fig.~\ref{fig:ttv}. The median values of the resultant probability distributions for $T_{C, 4}$ and $T_{C, 5}$, as well as the 68\%, 95\%, and 99\% confidence intervals, are provided in Table~\ref{tab:tcpredict}.

\begin{table}
    \centering
    \begin{tabular}{c | c | c }
        & $T_{C, 4}$ & $T_{C, 5}$  \\
        & (BJD TDB - 2450000)& (BJD TDB - 2450000) \\
        \hline
        Median & 9355.087 & 9897.078\\
        68\% & 9355.064 - 9355.118 & 9897.018 - 9897.192 \\
        95\% & 9355.032 - 9355.156 & 9896.910 - 9897.293 \\
        99\% & 9355.020 - 9355.205 & 9896.884 - 9897.455
    \end{tabular}
    \caption{Confidence intervals for the predictions of $T_{C, 4}$ and $T_{C, 5}$.}
    \label{tab:tcpredict}
\end{table}

\section{Conclusions}\label{sec:conc}
We detected a transit of \Nplanet\ on 2019 November 24 and 25, the first ever ground-based transit detection for this planet, primarily with data obtained using NGTS. As this is just the third ever transit of \Nplanet\ detected, we confirm the \Nperiodshort\,day orbital period of this planet.  This long orbital period makes \Nplanet\ the longest period planet to have a transit detected from the ground. The long transit duration (19\,hours) makes detecting transits of \Nplanet\ challenging and makes photometric efforts with increased longitudinal coverage useful for recovering such transits. On the other hand, the fact that many of the observations taken for this particular effort were not able to be used to constrain the transit parameters highlights the value of photometric facilities like NGTS, when it comes to obtaining high precision photometry of bright stars like \Nstar.

We found the transit to arrive \Ndtcshort\,hours earlier than predicted from a linear extrapolation of the \ktwo\ transits \citep{Vanderburg16hip41378, Becker19hip41378} and extensive spectroscopic monitoring \citep{Santerne19hip41378}. Using analytic formulae for TTV signals from \citet{lithwick12ttv}, we have shown that the TTV observed for \Nplanet\ can be explained solely through the interaction between \Nplanete\ and \Nplanet. Therefore, the presence of additional planets in the \Nstar\ system is not required to explain our observations. However our observations do not rule out the presence of additional planets. We predict that the next transit of \Nplanet\ will be centred on BJD \Ntcfourshort\ with a 99\% confidence interval of \Ntcfourwindow\,hours (2021 May 20), and the following transit will occur around BJD \Ntcfiveshort\ with a 99\% confidence interval of \Ntcfivewindow\,hours (2022 Nov 13).

Both $P_{\rm ttv}$ and $V$ depend strongly on the ratio of $P_f / P_e$. As $P_f$ is well constrained by the available transits of \Nplanet, this results in the amplitude and period of the predicted TTV signal depending strongly on the orbital period of \Nplanete. The current large uncertainty on $P_e$ combined with the shallow transit depth of just 1.5\,mmag mean that detecting another transit of \Nplanete, especially with ground-based facilities, will be tough. However, doing so would greatly refine the predictions of the TTV signal of \Nplanet. Additionally, detecting the upcoming transits of \Nplanet\ will allow us to greatly refine the ephemeris of \Nplanete, due to the constraints that the improved measurements of $V$ and $P_{\rm ttv}$ will place on $P_e$. The next transit of \Nplanete\ is predicted to have a centre time of $T_C = 2459356 \pm 60$ (UT 2021 May 21). \Nplanetd\ is in near 4:3 mean-motion resonance with \Nplanete, and so based on our TTV models we expect \Nplanetd\ to also experience large TTVs. Therefore, detecting transits of \Nplanetd\ will further allow us to refine our TTV models and make better predictions for upcoming transits of \Nplanete. Using the sampling method from Section~\ref{sec:ttv_analysis}, we predict the next transit of \Nplanetd\ to have a transit mid-point of $T_C = 2459393.13 \pm 0.59$ (2021 June 27).

With 6 planets in the \Nstar\ system, many of which reside near to mean-motion resonances, there are likely to be lots of interactions between the planets. With only a small number of transit detections for each of the outer three planets, a full dynamical analysis of the system is beyond the scope of this work. However, this robust transit detection of \Nplanet\ by NGTS and the discovery of significant TTVs in the system demonstrates that with future photometric monitoring we will be able to characterise the amplitude and super-period of the TTV signal of \Nplanet. This monitoring will enable us to place independent constraints on the masses, orbital periods, and eccentricities of the planets in this remarkable system.

\section*{Acknowledgements}
Based on data collected by the NGTS project, the SPECULOOS network, the Asiago Observatory, the Las Cumbres Observatory global telescope network, the MuSCAT2 instrument, the PEST Observatory, the Lick Observatory, the Hazelwood Observatory, the Phillip's Academy Observatory, and the PROMPT-8 telescope at CTIO.  For full acknowledgments for these facilities we refer the reader to the facility papers cited in this Letter.

MRK is supported by the NSF Graduate Research Fellowship, grant No. DGE 1339067. DJA acknowledges support from the STFC via an Ernest Rutherford Fellowship (ST/R00384X/1). MNG acknowledges support from MIT's Kavli Institute as a Juan Carlos Torres Fellow.  NN is partly supported by JSPS KAKENHI Grant Numbers JP17H04574, JP18H01265 and JP18H05439, and JST PRESTO Grant Number JPMJPR1775, and a University Research Support Grant from the National Astronomical Observatory of Japan (NAOJ). JSJ acknowledges support by FONDECYT grant 1201371, and partial support from CONICYT project Basal AFB-170002.

\section*{Data Availability}
The NGTS, Asiago, SSO and SNO photometric data are available in the online supplementary material.




\bibliographystyle{mnras}
\bibliography{HIP41378} 








\bsp	
\label{lastpage}
\end{document}

%% file: commands.tex
\newcommand{\ktwo}{{\it K2}}

\newcommand{\spec}{{SPECULOOS}}

\newcommand{\mpl}{\mbox{M$_{p}$}}
\newcommand{\rpl}{\mbox{R$_{p}$}}

\newcommand{\rearth}{R$_{\oplus}$}
\newcommand{\mearth}{M$_{\oplus}$}
\newcommand{\gccc}{g\,cm$^{-3}$}

\newcommand{\tczero}{$T_{C, 0}$}

\newcommand{\tcthree}{$T_{C, 3}$}
\newcommand{\tcfour}{$T_{C, 4}$}
\newcommand{\tcfive}{$T_{C, 5}$}

%% file: properties.tex
\newcommand{\Nstar}{HIP\,41378}

\newcommand{\NVmagshort}{$8.93$}

\newcommand{\Nplanet}{HIP\,41378\,f}
\newcommand{\Nplanete}{HIP\,41378\,e}
\newcommand{\Nplanetd}{HIP\,41378\,d}

\newcommand{\Nperiodshort}{\mbox{$542.08$}} 

\newcommand{\Ntcthree}{\mbox{$2458813.0913\pm0.0046$}} 

\newcommand{\Nmass}{\mbox{$12\pm3$}}
\newcommand{\Nradius}{\mbox{$9.2\pm0.1$}}
\newcommand{\Ndensity}{\mbox{$0.09\pm0.02$}}

\newcommand{\Ndtc}{\mbox{$1.50\pm0.11$}} 
\newcommand{\Ndtcshort}{$1.50$} 
\newcommand{\NPmax}{$1084.159$}
\newcommand{\NLDA}{\mbox{$0.315\pm0.004$}}
\newcommand{\NLDB}{\mbox{$0.289\pm0.001$}}
\newcommand{\Ntcvan}{$2457186.91451$}

\newcommand{\Ntcfour}{\mbox{$2459355.087^{+0.031}_{-0.022}$}}
\newcommand{\Ntcfive}{\mbox{$2459897.078^{+0.114}_{-0.060}$}}
\newcommand{\Ntcfourshort}{$2459355.087$}
\newcommand{\Ntcfiveshort}{$2459897.078$}
\newcommand{\Ntcfourwindow}{$4.4$}  
\newcommand{\Ntcfivewindow}{$13.7$}
\newcommand{\nwalkers}{70}
\newcommand{\nsample}{10000}
\newcommand{\nburnin}{2500}

%% file: HIP41378f.bbl
\begin{thebibliography}{}
\makeatletter
\relax
\def\mn@urlcharsother{\let\do\@makeother \do\$\do\&\do\#\do\^\do\_\do\%\do\~}
\def\mn@doi{\begingroup\mn@urlcharsother \@ifnextchar [ {\mn@doi@}
  {\mn@doi@[]}}
\def\mn@doi@[#1]#2{\def\@tempa{#1}\ifx\@tempa\@empty \href
  {http://dx.doi.org/#2} {doi:#2}\else \href {http://dx.doi.org/#2} {#1}\fi
  \endgroup}
\def\mn@eprint#1#2{\mn@eprint@#1:#2::\@nil}
\def\mn@eprint@arXiv#1{\href {http://arxiv.org/abs/#1} {{\tt arXiv:#1}}}
\def\mn@eprint@dblp#1{\href {http://dblp.uni-trier.de/rec/bibtex/#1.xml}
  {dblp:#1}}
\def\mn@eprint@#1:#2:#3:#4\@nil{\def\@tempa {#1}\def\@tempb {#2}\def\@tempc
  {#3}\ifx \@tempc \@empty \let \@tempc \@tempb \let \@tempb \@tempa \fi \ifx
  \@tempb \@empty \def\@tempb {arXiv}\fi \@ifundefined
  {mn@eprint@\@tempb}{\@tempb:\@tempc}{\expandafter \expandafter \csname
  mn@eprint@\@tempb\endcsname \expandafter{\@tempc}}}

\bibitem[\protect\citeauthoryear{{Agol}, {Steffen}, {Sari}  \&
  {Clarkson}}{{Agol} et~al.}{2005}]{Agol05ttvs}
{Agol} E.,  {Steffen} J.,  {Sari} R.,   {Clarkson} W.,  2005, \mn@doi [\mnras]
  {10.1111/j.1365-2966.2005.08922.x}, \href
  {https://ui.adsabs.harvard.edu/abs/2005MNRAS.359..567A} {359, 567}

\bibitem[\protect\citeauthoryear{Barbary}{Barbary}{2016}]{Barbary16sep}
Barbary K.,  2016, \mn@doi [Journal of Open Source Software]
  {10.21105/joss.00058}, 1, 58

\bibitem[\protect\citeauthoryear{{Becker} et~al.,}{{Becker}
  et~al.}{2019}]{Becker19hip41378}
{Becker} J.~C.,  et~al., 2019, \mn@doi [\aj] {10.3847/1538-3881/aaf0a2}, \href
  {https://ui.adsabs.harvard.edu/abs/2019AJ....157...19B} {157, 19}

\bibitem[\protect\citeauthoryear{{Bertin} \& {Arnouts}}{{Bertin} \&
  {Arnouts}}{1996}]{sextractor1996}
{Bertin} E.,  {Arnouts} S.,  1996, \mn@doi [\aaps] {10.1051/aas:1996164}, \href
  {https://ui.adsabs.harvard.edu/abs/1996A&AS..117..393B} {117, 393}

\bibitem[\protect\citeauthoryear{{Bryant} et~al.,}{{Bryant}
  et~al.}{2020}]{bryant20multicam}
{Bryant} E.~M.,  et~al., 2020, \mn@doi [\mnras] {10.1093/mnras/staa1075}, \href
  {https://ui.adsabs.harvard.edu/abs/2020MNRAS.494.5872B} {494, 5872}

\bibitem[\protect\citeauthoryear{{Burdanov}, {Delrez}, {Gillon}  \&
  {Jehin}}{{Burdanov} et~al.}{2018}]{Burdanov2018}
{Burdanov} A.,  {Delrez} L.,  {Gillon} M.,   {Jehin} E.,  2018, {SPECULOOS
  Exoplanet Search and Its Prototype on TRAPPIST}.
p.~130, \mn@doi{10.1007/978-3-319-55333-7_130}

\bibitem[\protect\citeauthoryear{{Claret} \& {Bloemen}}{{Claret} \&
  {Bloemen}}{2011}]{claretbloemen2011LDcoeffs}
{Claret} A.,  {Bloemen} S.,  2011, \mn@doi [\aap]
  {10.1051/0004-6361/201116451}, \href
  {https://ui.adsabs.harvard.edu/abs/2011A&A...529A..75C} {529, A75}

\bibitem[\protect\citeauthoryear{{Collins} et~al.,}{{Collins}
  et~al.}{2018}]{collins18}
{Collins} K.~A.,  et~al., 2018, \mn@doi [\aj] {10.3847/1538-3881/aae582}, \href
  {https://ui.adsabs.harvard.edu/abs/2018AJ....156..234C} {156, 234}

\bibitem[\protect\citeauthoryear{{Delrez} et~al.,}{{Delrez}
  et~al.}{2018}]{Delrez18SPECULOOS}
{Delrez} L.,  et~al., 2018, in \procspie. p. 107001I (\mn@eprint {arXiv}
  {1806.11205}), \mn@doi{10.1117/12.2312475}

\bibitem[\protect\citeauthoryear{{Eastman}, {Siverd}  \& {Gaudi}}{{Eastman}
  et~al.}{2010}]{eastman2010}
{Eastman} J.,  {Siverd} R.,   {Gaudi} B.~S.,  2010, \mn@doi [\pasp]
  {10.1086/655938}, \href
  {https://ui.adsabs.harvard.edu/abs/2010PASP..122..935E} {122, 935}

\bibitem[\protect\citeauthoryear{{Foreman-Mackey}, {Hogg}, {Lang}  \&
  {Goodman}}{{Foreman-Mackey} et~al.}{2013}]{foremanmackey13}
{Foreman-Mackey} D.,  {Hogg} D.~W.,  {Lang} D.,   {Goodman} J.,  2013, \mn@doi
  [\pasp] {10.1086/670067}, \href
  {http://adsabs.harvard.edu/abs/2013PASP..125..306F} {125, 306}

\bibitem[\protect\citeauthoryear{{Gaia Collaboration} et~al.,}{{Gaia
  Collaboration} et~al.}{2016}]{GAIA}
{Gaia Collaboration} et~al., 2016, \mn@doi [\aap]
  {10.1051/0004-6361/201629512}, \href
  {http://adsabs.harvard.edu/abs/2016A%26A...595A...2G} {595, A2}

\bibitem[\protect\citeauthoryear{{Gaia Collaboration}, {Brown}, {Vallenari},
  {Prusti}, {de Bruijne}, {Babusiaux}  \& {Bailer-Jones}}{{Gaia Collaboration}
  et~al.}{2018}]{GAIA_DR2}
{Gaia Collaboration} {Brown} A.~G.~A.,  {Vallenari} A.,  {Prusti} T.,  {de
  Bruijne} J.~H.~J.,  {Babusiaux} C.,   {Bailer-Jones} C.~A.~L.,  2018,
  preprint, \href {http://adsabs.harvard.edu/abs/2018arXiv180409365G} {}
  (\mn@eprint {arXiv} {1804.09365})

\bibitem[\protect\citeauthoryear{{Gillon}}{{Gillon}}{2018}]{Gillon2018}
{Gillon} M.,  2018, \mn@doi [Nature Astronomy] {10.1038/s41550-018-0443-y},
  \href {https://ui.adsabs.harvard.edu/abs/2018NatAs...2..344G} {2, 344}

\bibitem[\protect\citeauthoryear{{Howell} et~al.,}{{Howell}
  et~al.}{2014}]{Howell14K2}
{Howell} S.~B.,  et~al., 2014, \mn@doi [\pasp] {10.1086/676406}, \href
  {https://ui.adsabs.harvard.edu/abs/2014PASP..126..398H} {126, 398}

\bibitem[\protect\citeauthoryear{Irwin et~al.,}{Irwin et~al.}{2004}]{Irwin2004}
Irwin M.~J.,  et~al., 2004, in Optimizing Scientific Return for Astronomy
  through Information Technologies. {SPIE}, \mn@doi{10.1117/12.551449}, \url
  {https://doi.org/10.1117/12.551449}

\bibitem[\protect\citeauthoryear{{Kawahara} \& {Masuda}}{{Kawahara} \&
  {Masuda}}{2019}]{KawaharaMasuda19longperiodKepler}
{Kawahara} H.,  {Masuda} K.,  2019, \mn@doi [\aj] {10.3847/1538-3881/ab18ab},
  \href {https://ui.adsabs.harvard.edu/abs/2019AJ....157..218K} {157, 218}

\bibitem[\protect\citeauthoryear{{Kempton} et~al.,}{{Kempton}
  et~al.}{2018}]{kempton2018tsm}
{Kempton} E. M.~R.,  et~al., 2018, \mn@doi [\pasp] {10.1088/1538-3873/aadf6f},
  \href {https://ui.adsabs.harvard.edu/abs/2018PASP..130k4401K} {130, 114401}

\bibitem[\protect\citeauthoryear{{Lithwick}, {Xie}  \& {Wu}}{{Lithwick}
  et~al.}{2012}]{lithwick12ttv}
{Lithwick} Y.,  {Xie} J.,   {Wu} Y.,  2012, \mn@doi [\apj]
  {10.1088/0004-637X/761/2/122}, \href
  {https://ui.adsabs.harvard.edu/abs/2012ApJ...761..122L} {761, 122}

\bibitem[\protect\citeauthoryear{{Lund} et~al.,}{{Lund}
  et~al.}{2019}]{Lund19hip41378asteroseismology}
{Lund} M.~N.,  et~al., 2019, \mn@doi [\aj] {10.3847/1538-3881/ab5280}, \href
  {https://ui.adsabs.harvard.edu/abs/2019AJ....158..248L} {158, 248}

\bibitem[\protect\citeauthoryear{{Murray} et~al.,}{{Murray}
  et~al.}{2020}]{Murray2020}
{Murray} C.~A.,  et~al., 2020, \mn@doi [\mnras] {10.1093/mnras/staa1283}, \href
  {https://ui.adsabs.harvard.edu/abs/2020MNRAS.495.2446M} {495, 2446}

\bibitem[\protect\citeauthoryear{{Nascimbeni}, {Piotto}, {Bedin}  \&
  {Damasso}}{{Nascimbeni} et~al.}{2011}]{nascimbeni2011a}
{Nascimbeni} V.,  {Piotto} G.,  {Bedin} L.~R.,   {Damasso} M.,  2011, \mn@doi
  [\aap] {10.1051/0004-6361/201015199}, \href
  {https://ui.adsabs.harvard.edu/abs/2011A&A...527A..85N} {527, A85}

\bibitem[\protect\citeauthoryear{{Santerne} et~al.,}{{Santerne}
  et~al.}{2019}]{Santerne19hip41378}
{Santerne} A.,  et~al., 2019, arXiv e-prints, \href
  {https://ui.adsabs.harvard.edu/abs/2019arXiv191107355S} {p. arXiv:1911.07355}

\bibitem[\protect\citeauthoryear{{Sebastian} et~al.,}{{Sebastian}
  et~al.}{2021}]{Sebastian2021}
{Sebastian} D.,  et~al., 2021, \mn@doi [\aap] {10.1051/0004-6361/202038827},
  \href {https://ui.adsabs.harvard.edu/abs/2021A&A...645A.100S} {645, A100}

\bibitem[\protect\citeauthoryear{{Smith} et~al.,}{{Smith}
  et~al.}{2020}]{smith20multicam}
{Smith} A. M.~S.,  et~al., 2020, arXiv e-prints, \href
  {https://ui.adsabs.harvard.edu/abs/2020arXiv200205591S} {p. arXiv:2002.05591}

\bibitem[\protect\citeauthoryear{{Vanderburg} et~al.,}{{Vanderburg}
  et~al.}{2016}]{Vanderburg16hip41378}
{Vanderburg} A.,  et~al., 2016, \mn@doi [\apjl] {10.3847/2041-8205/827/1/L10},
  \href {https://ui.adsabs.harvard.edu/abs/2016ApJ...827L..10V} {827, L10}

\bibitem[\protect\citeauthoryear{{Wheatley} et~al.,}{{Wheatley}
  et~al.}{2018}]{Wheatley18NGTS}
{Wheatley} P.~J.,  et~al., 2018, \mn@doi [\mnras] {10.1093/mnras/stx2836},
  \href {https://ui.adsabs.harvard.edu/abs/2018MNRAS.475.4476W} {475, 4476}

\makeatother
\end{thebibliography}
